\documentstyle[12pt]{article}
\newcommand{\ZZ}{{\rm \kern 0.275em Z \kern -0.92em Z}\;}
\newcommand{\be}{\begin{equation}}
\newcommand{\ee}{\end{equation}}

\begin{document}

\begin{center}
{\Large Non-Boltzmann Statistics as an Alternative to Holography}
\\
\vspace*{1.2cm}
 {\large Marcelo Botta
Cantcheff$^{\dag,}$$^{\ddag}$\footnote{botta@cbpf.br} and Jose A.C.
Nogales$^{\ddag,}$$^{\star,}$\footnote{jnogales@fiumsa.edu.bo,
jnogales@cbpf.br}}
\\
\vspace*{0.7cm} $^{\dag}$Instituto de F\'{\i}sica Te\'orica
(IFT/UNESP)

Rua Pamplona 145 - Bela Vista
01405-900- Sao Paulo - SP - Brazil.

\vspace{3mm}

$^{\ddag}$Centro Brasileiro de Pesquisas F\'{\i}sicas (CBPF),\\
Departamento de Teoria de Campos e Part\'{\i}culas (DCP),\\ Rua
Dr. Xavier
Sigaud 150, 22290-180, Rio de Janeiro, RJ, Brasil.\\
\vspace{3mm}

\vspace*{0.3cm} $^{\star}$ Instituto de F\'{\i}sica, Facultad de
Ciencias Puras y Naturales,\\ Calle 27 Campus Universitario,
Universidad Mayor de San Andr\'{e}s,\\ Casilla 8635, La Paz, Bolivia
\end{center}

\begin{abstract}
An intriguing question related to black hole thermodynamics is
that the entropy of a region shall scale as the area rather than
the volume. In this essay we propose that the microscopical
degrees of freedom contained in a given region of space, are
statistically related in  such a way that obey a non-standard
statistics, in which case an holographic hypothesis would be not
needed. This could provide us with some insight about the nature
of degrees of freedom of the geometry and/or the way in which
gravitation plays a role in the statistic correlation between the
degrees of freedom of a system.

\end{abstract}




\newpage
We start this note by making the following crucial observation:
when gravitational interaction of a given system is considered, it
is impossible to isolate two subsystems (unless they are
infinitely separated) \footnote{In contrast with ordinary
thermodynamic systems.}, so, to define statistical independence of
two subsystems is actually problematic in this context. On the
other hand, it is known that for systems whose microscopical
components cannot be considered statistically independent (and in
particular, those being highly correlated), it shall be considered
a non-standard statistics (non-additive or
non-extensive)\cite{correl}. Consequently, we can expect a
deviation of the standard statistics when gravity is considered an
important part of a system.

In the last two decades, generalizations of the Boltzmann-Gibbs
thermodynamics have been extensively explored \cite{Qent}.

In an ideal system in thermodynamic equilibrium, if the intensive
variables are kept fixed and the side of the system is doubled,
the extensive variables should then also double. This is not true
for black holes
\cite{Bekenstein.1972,Carter.1973,Hawking.1975,Jacobson.1995}.

 From the thermodynamic point of view, we mean
macroscopic \cite{botta}, the entropy of black holes is
proportional to the area of the event horizon
$$S\propto A .$$ We are going to analyze
 the statistical behavior of the degrees of freedom that compose
this type of systems, in order to have consistency with the
relation above.

Let us explain more carefully the difference between additivity
and extensivity \cite{Tsallis.2004}. Consider a physical quantity
$ W(i)$ related with the subsystem $i$. For $N$ of such
subsystems, $W$ is additive if we have:  \be W\left( N
\right)=\sum_{i=1}^{N}W\left(i\right)\;,\label{e1}\ee where
$W(N)\equiv W\left(\{i \}_1^{N}\right)$.  Supposing that all
subsystems are equal, \be W(N)=N W(1)\;.\label{e2}\ee In its turn,
extensivity is defined by:
\be\lim_{N\rightarrow\infty}\frac{|W(N)|}{N}<\infty
\;.\label{e3}\ee The physical quantities that do not fulfill this
property are called non-extensive. We note that all quantities
which are additive are also extensive, in fact \be
\lim_{N\rightarrow\infty}\frac{|W(N)|}{N}= {|\bar W|}\;, \ee where
the mean value, $|{\bar W}|$, is finite since $W(i)< \infty \,\,
\forall i$. In particular if the subsystems are equal, we get
$\lim_{N\rightarrow\infty}\frac{|W(N)|}{N}= W(1)$.

\section{Blak Hole thermodynamics and the holographic interpretation}

Roughly speaking, the argument which leads to the holographic principle is that if one 
assumes an uniform distribution of the gravitational degrees of freedom in a spatial 
region of volume $V$, or in other words, if, as suggested by quantum gravity, the 
space-time is thought as filled of quantum geometric atoms like
wormholes or more complex structures of Planck size
$\ell_{p}$ more or less uniformly distributed, then the number of microstates is $\propto 
M^V $, where $M$ is the number of states reachable by each microscopical component. Then, 
following the Boltzmann (extensive) statistics, one obtains that entropy $S$ must scale 
as the volume $V$ which would lead to a contradiction with the Bekenstein-Hawking law for 
the black hole entropy, $S=A/4$ in Planck´s units \cite{Bekenstein.1972,Hawking.1975}.
 So, the holographic principle suggested by 
t´Hooft \cite{Hooft.1993} essentially claims that all these degrees of freedom may be mapped to 
the boundary of that region, and consequently the corresponding entropy would 
scale as the area.

As we can notice, an important part of this argument is the extensive Boltzmann-Gibbs 
statistics, and our proposal is precisely based on alternatives to this, in order to match 
consistency with the area law and to render the holographic
viewpoint not necessary \footnote{In this approach, we are not
worried with the aspects concerning the information paradox which,
in principle, could be avoided by the holographic principle.}.

\section{Non-standard statistics and consistency with the area law.}

Now, we keep the assumption of uniformity in the distribution of
the physical degrees of freedom in a spatial region of volume $V$,
and drop out the holographic point of view about the possibility
of mapping them to the boundary.

According to this, if we suppose $N$ identical degrees of freedom
uniformly distributed in $V$, we have $N \propto V $. Following
the area law, we must have $S(N)\propto A  \propto N^{2/3}$. Thus,
our main result is: \be S(N)\sim N^{2/3} S(1) \;,\label{e4}\ee
which differs from (\ref{e2}). Substituting this into (\ref{e3})
we get \be \lim_{N\rightarrow\infty}\frac{|S(N)|}{N}=0  \; ,\ee
which shows the extensivity of the entropy.

Our conclusion is that the statistics that must be considered in a
system where the gravitation plays an important role (as a Black
Hole) is non-standard, in particular the entropy {\it is a
non-additive but extensive } quantity. The law which rules the non
additivity is given by (\ref{e4}).

\section{Correlated Systems and Tsallis
Statistics.}

Finally, we  are going to discuss here a possible realization of
the framework described above. As discussed in \cite{correl}, for
certain systems where the hypothesis of probabilistic independence
is not applicable, the number of states is \be \Gamma \sim
N^\gamma \;\;\; (\gamma >0 \, , \, N \gg 1 ). \label{e5}\ee
 On the other hand, the corresponding
expression for the Tsallis entropy is \be\label{qent} S = k
\frac{\Gamma^{1-q}-1}{1-q}\, , \ee where $q<1$ is a dimensionless
parameter.

If the degrees of freedom, {\it geometric or not}, are correlated
by gravitational interaction such that (\ref{e5}) holds, we obtain
that \be S \sim N^{(1-q) \gamma} \,\,\, ( N \gg 1) \,\, .\ee Thus,
by requiring consistency with (\ref{e4}), we obtain: \be (1-q)
\gamma = 2/3 . \label{qp}\ee As a particular example, if we have a
sort of Boolean system where each (distinguishable) degree of
freedom may assume the states $s_i = 0,1 \;\; , \; i=1,..., N $
\cite{Hooft.1993}, and the constraint \be \sum_{i=1}^{N} s_i = 1
\,\, \label{c}\ee is also assumed, then, the possible states of
the system are given by $s_i = \delta_{i k}$, for some $k \, (1
\leq k \leq N)$. Then, we have $\Gamma = N$. Substituting this
into (\ref{qp}), as a result one has: $\gamma \sim 1 \Rightarrow
q=1/3$.


On the other hand, notice that the $q$-statistics may be
continuously extended to $q>1$ values, and expression (\ref{qent})
becomes \be S = k \frac{\Gamma^{q-1}-1}{q-1}\, . \ee In this case,
one can verify a remarkable coincidence with a conjecture due to
Tsallis et al \cite{conjetura,Tsallis.2004}, about statistics of
systems governed by a potential decreasing with the inverse square
of the distance (as is the case of gravity). By using arguments
quite different from here, they argued that $q$ should be $ 5/3$.
But, on the other hand, we obtain precisely this value ($q \sim
5/3$) for the simplified model we are considering above (with
$\gamma \sim 1$).

\vspace{1.3cm}

We conclude this work
 by stressing that also the problems
associated with the black hole thermodynamical instability could
be by-passed by considering certain non-standard statistics
\cite{nos}.

\vspace{4cm}


  We would like to thank to CNPq for the invaluable financial
  help. Special thanks are due to A.L.M.A. Nogueira and J. A. Helayel by reading the
  manuscript.


\end{document}